\documentclass[twocolumn,11pt]{IEEEtran_v15}
\usepackage{amsmath}
\usepackage{psfrag}
\usepackage{graphicx}
\usepackage{subfigure}
\usepackage{cite}

\newlength{\figwidth}
\newcommand{\ms}{\mathrm{ms}}
\newcommand{\bps}{\mathrm{bps}}
\newcommand{\TRTT}{T_{\textrm{\tiny RTT}}}
\newcommand{\hT}{K_0}
\newcommand{\ns}{\texttt{ns}}
\newcommand{\Sn}{\left\{S\right\}_n}

\begin{document}
\title{On the Prospects of Chaos Aware Traffic Modeling}
\author{
  \large
  \begin{tabular}[t]{c c c c}
    A. Fekete,\!\textsuperscript{1, 2} &
    M. Mar\'odi,\!\textsuperscript{1, 3} & and & G. Vattay\textsuperscript{1, 3}
  \end{tabular}\\[2mm]\normalsize
  \begin{minipage}[t]{0.5\textwidth}
    \begin{tabbing}
      \textsuperscript{1} \= Department of Physics of Complex Systems\\
      \> E\"otv\"os University, Budapest, Hungary\\
      \textsuperscript{2} \> Ericsson Hungary Ltd., Budapest, Hungary\\
      \textsuperscript{3} \> Communication Networks Laboratory\\
      \> E\"otv\"os University, Budapest, Hungary
    \end{tabbing}
  \end{minipage}
  }
\date{}
\maketitle

\begin{abstract}
  In this paper the chaotic properties of the TCP congestion avoidance
  mechanism are investigated. The analysis focuses on the origin of
  the complex behavior appearing in deterministic TCP/IP networks.
  From the traffic modeling point of view the understanding of the
  mechanism generating chaos is essential, since present models are
  unable to cope with this phenomena.
  
  Using the basic tools of chaos theory in our study, the main
  characteristics of chaotic dynamics are revealed. The dynamics of
  packet loss events is studied by a simple symbolic description. The
  cellular structure of the phase space of congestion windows is
  shown. This implies periodic behavior for large time scales.
 
  Chaotic behavior in short time scales and periodicity for larger
  times makes it necessary to develop models that account for both.
  Thus a simple model that describes the congestion window dynamics
  according to fluid equations, but handles the packet loss events
  separately is introduced. This model can reproduce the basic
  features observed in realistic packet level simulations.
\end{abstract}


\section{Introduction}

Recently Veres and Boda \cite{Veres_B} have demonstrated that TCP
congestion control can be chaotic in certain circumstances. Chaos in
practice means that TCPs influencing each other in a computer network
can produce highly complex behavior in time which is sensitive to
small perturbations, yet the equations describing it are deterministic
and simple.  Understanding the mechanism and equations that produce
chaos in a TCP/IP network is crucial in traffic modeling. If we can
identify the details of the mechanism then it is possible to build new
kinds of network traffic models where the factors important from the
point of view of dynamics and chaos are kept while many unimportant
factors are reduced or simplified. Current TCP models are either too
simplistic by assuming periodic behavior or they are entirely
stochastic disregarding small details of a given network which might
alter the dynamics completely and even change its statistical
properties.  Stochastic models can also break the temporal and spatial
(topological) structure of correlations existing among TCPs in an
extended network and are not able to give a correct account of
possible long range dependence born between far away TCPs
\cite{Veres_K_M_V}. On the other hand, new chaos aware TCP and network
models can preserve correlations and show the same level of complexity
as it is observed in real network traffic. These features might turn
out to be unavoidable when building reliable models of large networks
where packet level simulation requires astronomical computer resources
and/or simulation time.  Moreover, chaos theory itself provides new
tools to quantify this complexity. The complexity of a real network
traffic or its packet level simulation and the complexity of a traffic
model can be compared.  Despite their inherent simplicity, these tools
have never been used in networking and they might open new prospects
in verifying TCP and network models or can even characterize the
actual network state.

We think that all these issues are so important, that we should answer
at least the basic open questions concerning the chaotic state of the
TCP congestion control mechanism. For example we still do not know
what causes TCP chaos exactly. Accordingly we do not know how generic
its appearance is. For example, it has been argued in
Ref.~\cite{Guo_C_M} that in the chaotic examples shown in
Ref.~\cite{Veres_B} the packet loss probability is considerably higher
than 1\%. It has been shown \cite{Guo_C_M,Figue_L_M_T,Fekete_V} that
in this case the exponential backoff mechanism plays an important role
and can be responsible for the complex congestion window dynamics. We
are going to show that chaos is not a consequence of high network
congestion or loss and that TCPs operating in congestion avoidance
mode, never entering into a backoff state show chaos. In other words
chaos is the generic behavior of many TCP systems and periodicity and
synchronization is rather exceptional.

\begin{figure}[htbp]
  \begin{center}
    \psfrag{Buffer}[c][c][1]{Buffer}
    \psfrag{Internet}[c][c][1]{Network} \psfrag{B}[c][c][1]{$B$}
    \psfrag{D0}[c][c][1]{$T_0$} \psfrag{D1}[c][c][1]{$T_1$}
    \psfrag{D2}[c][c][1]{$T_2$} \psfrag{D3}[c][c][1]{$T_3$}
    \psfrag{C0}[c][c][1]{$C_0$} \psfrag{C1}[c][c][1]{$C_1$}
    \psfrag{C2}[c][c][1]{$C_2$} \psfrag{C3}[c][c][1]{$C_3$}
    \resizebox{\figwidth}{!}{\includegraphics{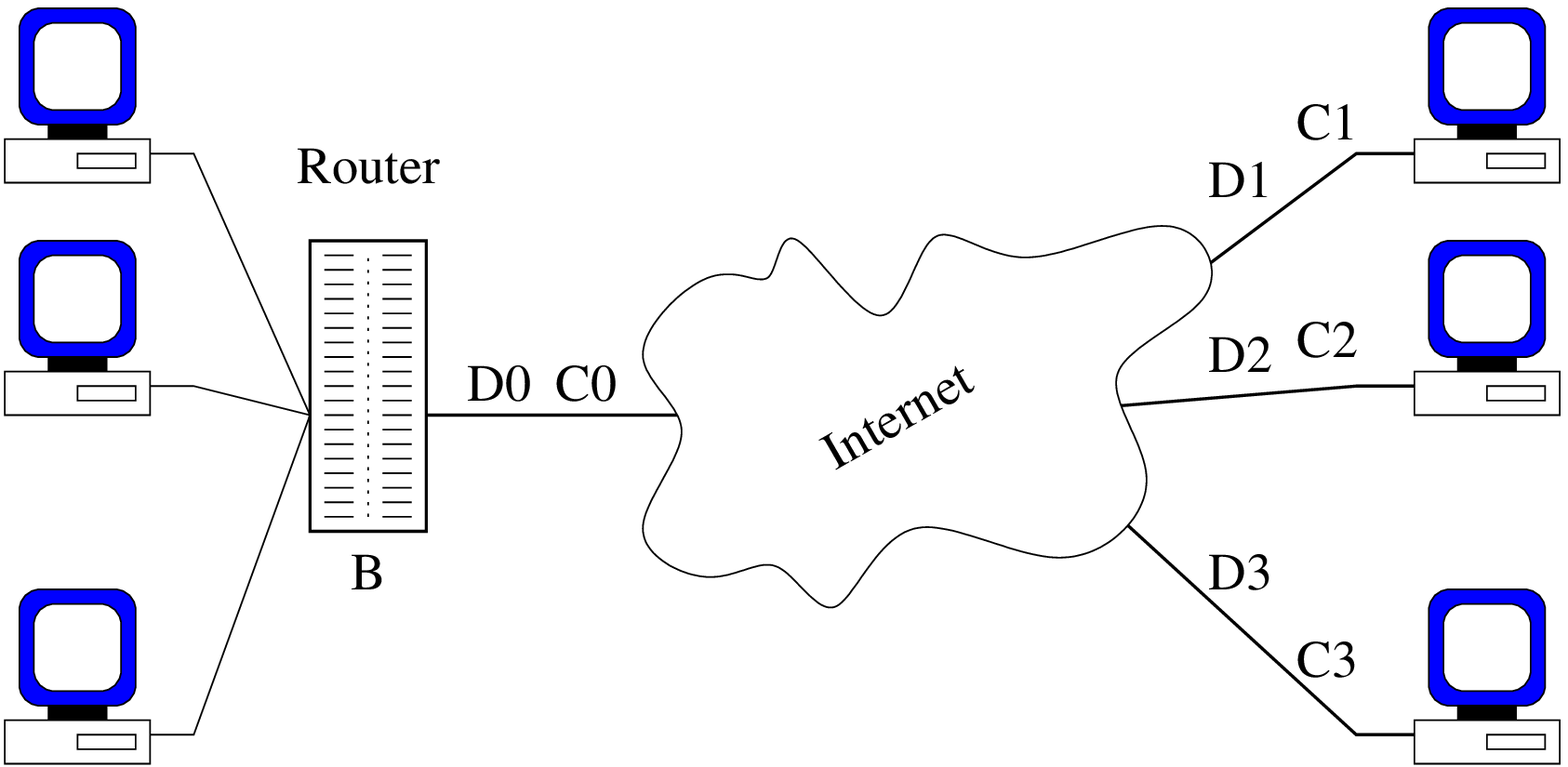}}
    \caption{Investigated network model: There are three TCP flows
      sharing a common buffer that can store $B$ packets and a common
      line with delay $D_0$ and speed $C_0$. Then the common link
      splits into three different lines with different delays and
      speeds. In the actual simulations $B=100$, $T_0=400\,\ms$,
      $T_1=100\,\ms$, $T_2=150\,\ms$, $T_3=200\,\ms$, $C_0=10^6\,\bps$
      and $C_1=C_2=C_3=10^7\,\bps$ has been chosen.}
    \label{fig:model}
  \end{center}
\end{figure}

Also we have to ask how we can distinguish TCP chaos from
stochasticity and do we gain anything by doing that. To point out the
main weakness of stochastic models and to call for chaos aware models
we would like to show that current stochastic TCP models are even
unable to predict the traffic in a simple scenario like the one shown
in Fig.~\ref{fig:model} and investigated throughout this paper.
In this setup three TCP flows sharing a common buffer that can store
$B$ packets and a common line with delay $T_0$ and speed $C_0$, then
splitting into three different lines with different delays and speeds.
In the actual simulations $B=100$, $T_0=400\,\ms$, $T_1=100\,\ms$,
$T_2=150\,\ms$, $T_3=200\,\ms$, $T_0=10^6\,\bps$ and
$C_1=C_2=C_3=10^7\,\bps$ has been chosen. The congestion windows of
the competing TCPs are not limited by the senders or by the receivers.
The injected TCP packets can be lost only at the bottleneck buffer,
there are no random losses on the links or in other buffers.  It
follows that the traffic is controlled by strict deterministic rules,
that is by the TCP Reno algorithm \cite{RFC2001}. Numerical
simulations were carried out by Network Simulator (\ns) version 2b5
\cite{ns}.

\begin{figure}[htbp]
  \begin{center}
    \psfrag{t}[c][c][1]{$t$}
    \psfrag{cwnd}[c][c][1]{Congestion window}
    \psfrag{perturbation}[c][c][1]{perturbation}
    \psfrag{cwnd_1}[r][r][1]{TCP$_1$}
    \psfrag{cwnd_2}[r][r][1]{TCP$_2$}
    \psfrag{cwnd_3}[r][r][1]{TCP$_3$}
    \subfigure[]{
      \resizebox{\figwidth}{!}{\includegraphics{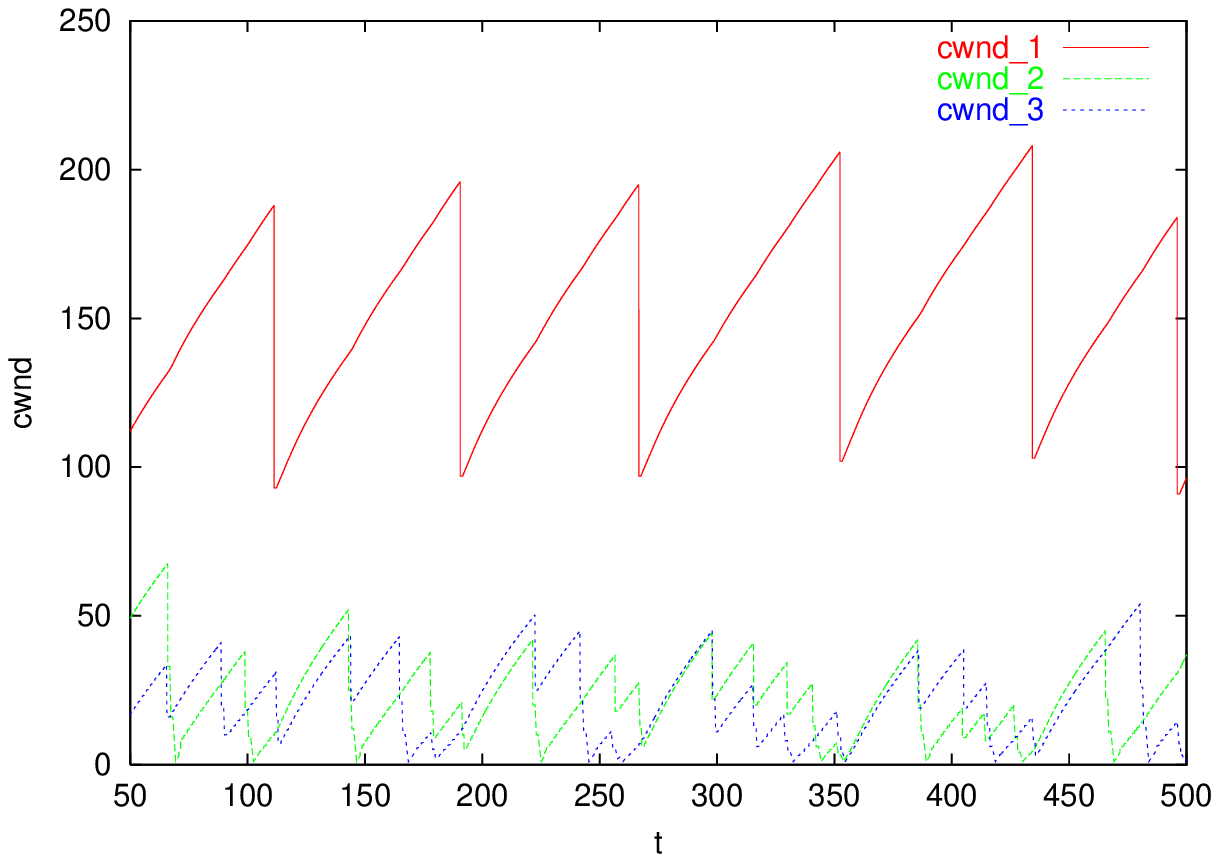}}
      \label{subfig:cwnd_nopert}
      }
    \subfigure[]{
      \resizebox{\figwidth}{!}{\includegraphics{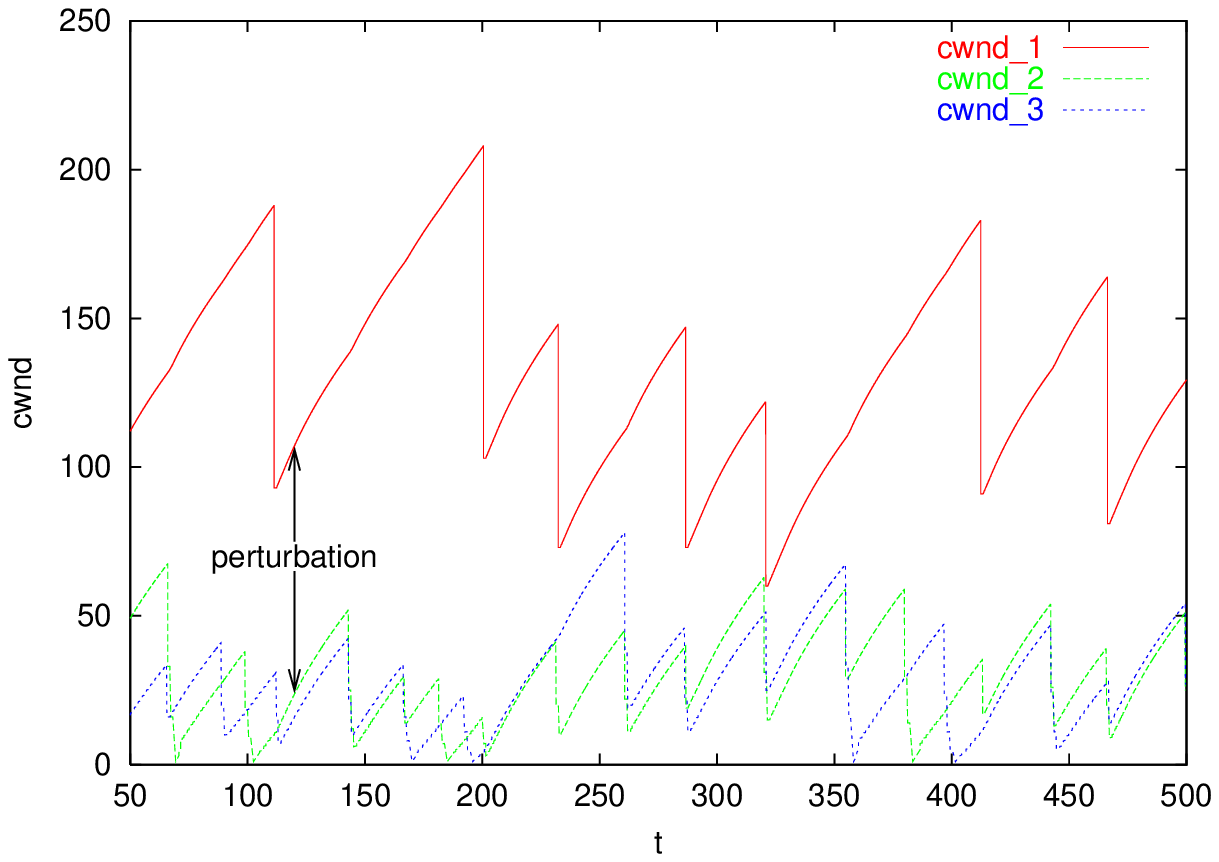}}
      \label{subfig:cwnd_pert}
      }
    \caption{(a) A typical part of the congestion window time
      series is shown for the simulation of the scenario of
      Fig.\ref{fig:model}. (b) The effect of a small perturbation
      at $t=120\,\textrm{s}$ on the congestion window development.}
    \label{fig:cwnd_sample}
  \end{center}
\end{figure}

Based on ideas brought from stochastic TCP modeling a common belief
is that TCP is biased against long round trip time connections and the
throughputs are proportional to $\sim 1/\TRTT^2$. This assumption has
been proven to be very good in the presence of random elements in the
simulation \cite{Zhang_Q_K}. In reality, shown in
Fig.~\ref{subfig:cwnd_nopert} it can be observed that the congestion
window corresponding to the {\em largest} round trip time is
significantly larger than the others. This TCP obtains \emph{unfairly}
higher throughput than the other two. The packet loss rate of the
preferred TCP ($\sim5\cdot10^{-5}$) is also an order of magnitude less
than that of the suppressed ones ($\sim7\cdot10^{-4}$). The ``winner''
TCP behaves seemingly periodically, while the ``losers'' seem to be
erratic.
Obviously, there is something here which is missed completely by the
stochastic model.  Next, we show that this simple scenario is already
chaotic and the deterministic nature of packet losses cannot be
disregarded if we would like to build models that correctly predict
the temporal behavior of congestion windows.

\section{The TCP butterfly effect}

The complete state of a TCP can be given by a number of internal
variables at any moment \cite{RFC2001}. Such variables are the
congestion window, the slow start threshold, the retransmission
timeout, the backoff counter, the duplicate ACK counter, and so on.
However, during the optimal operation of TCP, in congestion avoidance
mode, a single variable can be selected which controls almost
completely the behavior of the TCP: that is the congestion window.
This variable has also practical importance, since it limits the
maximum number of unacknowledged packets sent by a TCP into the
network.

To demonstrate how \emph{sensitive} this system can be for small
perturbations the same simulation has been run for $120\,\mathrm{s}$
and then a perturbation of $\delta w_i(0)=0.01$ has been added to all
the congestion window values at $t=120\,\mathrm{s}$. The result is
shown in Fig.~\ref{subfig:cwnd_pert}. The congestion windows remain
unchanged until $t=120\,\mathrm{s}$. Then the difference between the
congestion windows of the two simulations $|\delta\mathbf{w}(\tau)|$
remains the same ($\sim 0.01$) until the first packet drop event. Then
one of the underprivileged TCPs, whose packet has been lost, halves
its window. As a result, the distance between the original and the
perturbed trajectories grows about an order of magnitude, since the
owner of the lost packet differs in the original and in the perturbed
simulations. At this point it seems that the dominant TCP is not
affected at all. Finally, around $t=240\,\mathrm{s}$ the time
evolution of the dominant TCP diverges completely from the original
trace due to a permutation of packets, resulting in a loss event for
the dominant TCP. As we can see the rest of the simulation differs
from the original one. Such sensitivity against small perturbations is
called the butterfly effect in chaos theory and gives us the first
clue that this system operating in congestion avoidance is actually
chaotic. Next we introduce a few basic tools which help us to
characterize the chaotic state.

\section{Characterizing chaos}
\label{sec:Characterizing}

\begin{figure}[htbp]
  \begin{center}
    \psfrag{w_1}[c][c][1]{$w_1$}
    \psfrag{w_2}[c][c][1]{$w_2$}
    \psfrag{w_3}[c][c][1]{$w_3$}
    \resizebox{\figwidth}{!}{\includegraphics{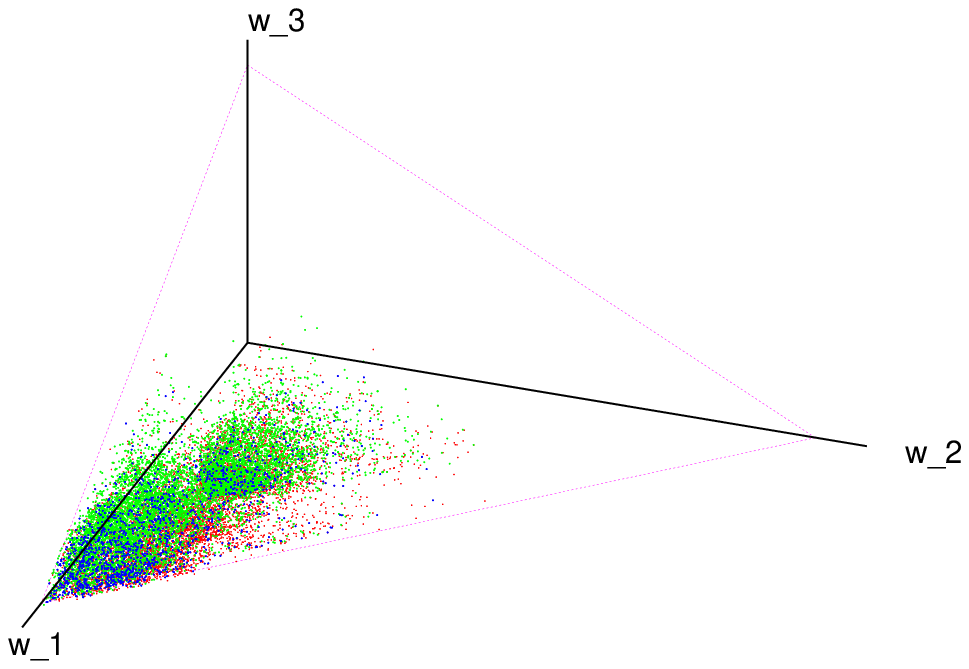}}
    \caption{Poincar\'e section of the phase space. Congestion window
      values for the three TCPs at packet loss times. Colours depend
      on which TCP lost a packet. The depicted (pink) surface is where
      the sum of congestion windows is approximately equal with the
      maximum number of packets which can travel on the lines or stay
      in the buffer.}
    \label{fig:pss}
  \end{center}
\end{figure}

One of the most basic tools of chaos theory in visualizing the
dynamics is the Poincar\'e surface of section. Instead of looking at
the continuous time evolution of trajectories one can select a surface
in the phase space and watch only when the trajectories cross that
surface. In case of TCP congestion window trajectories the evolution
between two packet loss events is fairly simple. All the interesting
things happen at the times of packet losses. Therefore the values of
congestion windows taken at the moments of packet losses of any of the
TCPs is a natural choice for a surface of section in general. In
Fig.~\ref{fig:pss} this surface of section is shown for our system.
One can see that the congestion window triplets $(w_1,w_2,w_3)$ taken
at times of packet losses approximately form a two dimensional surface
within the three dimensional congestion window phase space. It is easy
to understand why we get such a surface: packet losses occur when the
buffer is full. In the scenario of Fig.~\ref{fig:model} the packets
first fill the lines denoted by 0, 1, 2 and 3. The number of packets
which can travel on these lines is given by the bandwidth delay
products divided by the packet size $C_0T_i/P$, where the packet size
in our simulations was $512$ bytes. The maximum number of packets on the
lines and in the buffer might be approximated by
$Q=B+C_0T_0/P+\frac{1}{3}C_0(T_1+T_2+T_3)/P$, as all packets go through
link $0$, while each packet should choose either one of the three lines
$1$, $2$ or $3$. The sum of congestion windows $W=\sum_i w_i$ is
approximately the number of packets in the network and packet loss
occurs approximately when $W=Q$. This equation defines the ``surface
of loss'' inside the window phase space. This surface is also
indicated in Fig.~\ref{fig:pss}.

In chaotic systems the attractor is often a fractal object. Fractals
are statistically self-similar geometric objects that might be
characterized by suitably defined non-integer valued dimensions. For
ordinary fractals this dimension is less then the Eucledian embedding
dimension $D$ of the object \cite{Vicsek}. The fractal dimension of
the points on the surface of section can be measured. To do this we
can project the points onto a suitable surface. The points were
projected onto the $\sum w_i=\mathit{const.}$ surface and the fractal
dimension of this two dimensional projection was measured. A usual
method for measuring the fractal dimension is when a grid of cells of
size $\epsilon$ is put on the object, and the number of non-empty
cells $N(\epsilon)$ is counted. The \emph{box counting dimension} is
given by
\begin{equation}
  D_0=-\lim_{\epsilon\rightarrow 0}\frac{\log N(\epsilon)}{\log \epsilon}.
\end{equation}  

\begin{figure}[htbp]
  \begin{center}
    \psfrag{log10(l)}[c][c][1]{$\log_{10}(\epsilon)$}
    \psfrag{log10(N(l))}[c][c][1]{$\log_{10}(N(\epsilon))$}
    \psfrag{log10(sum(p_i^2))}[c][c][1]{$\log_{10}(\sum p_i^2)$}
    \psfrag{1.69 pm 0.02}[l][l][1]{$D_0=1.69\pm 0.02$}
    \psfrag{1.70 pm 0.05}[r][r][1]{$1.70\pm 0.05$}
    \resizebox{\figwidth}{!}{\includegraphics{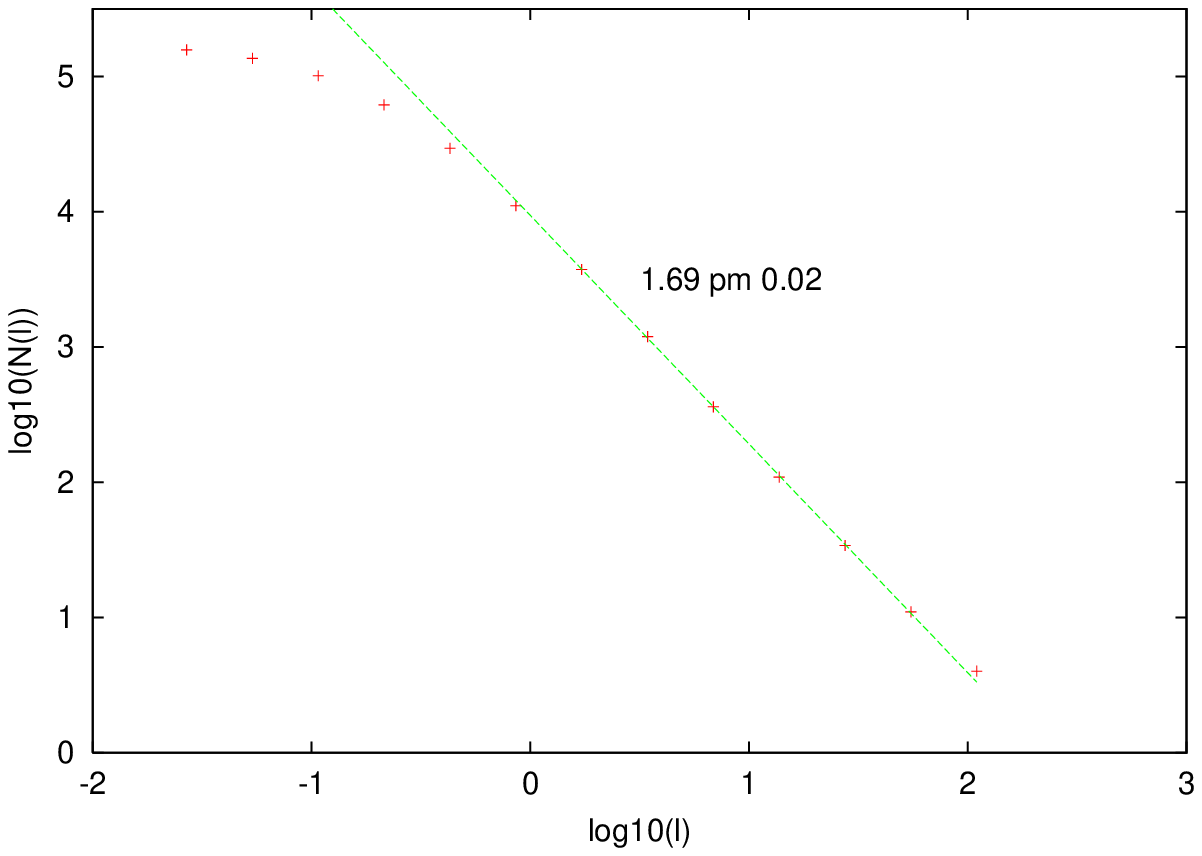}}
    \caption{Fractal dimension of the two dimensional projection of
      the surface of section of Fig.~\ref{fig:pss}.}
    \label{fig:fracdim_sim}
  \end{center}
\end{figure}

In Fig.~\ref{fig:fracdim_sim} it can
be seen that the points on the surface form a non-trivial fractal
with box counting dimension $D_0=1.69 \pm 0.02$.

Now our qualitative picture of TCP dynamics in congestion avoidance
mode can be summarized as follows. Congestion windows steadily grow
between packet losses. This process can usually be well
approximated \cite{Ott_K_M,Misra_O} with fluid equations of the type
\begin{equation}
  \frac{dw_i(t)}{dt}=\frac{1}{T_{\textrm{\tiny RTT},i}(\mathbf{w})},
  \label{eq:fluids}
\end{equation}
where $T_{\textrm{\tiny RTT},i}(\mathbf{w})$ is the round trip time of
the $i$th TCP. Round trip times can also be approximated as functions
of the congestion windows and then the resulting differential
equations (\ref{eq:fluids}) can be solved self-consistently. The sum
of congestion windows also grows steadily and the congestion windows
cut the surface of loss at some point. Then one of the congestion
windows is halved according to the TCP algorithm and the position of
the point in the phase space drops below the loss surface and the
process starts again.

\section{Symbolic description}
\label{sec:Symbols}

One can see that the dynamical process between packet losses described
above is relatively simple. The complicated fractal structure of the
attractor and the sensitivity for perturbations should come from the
fine details of the packet loss process.  Each time the congestion
window trajectory crosses the loss surface a packet loss event happens
in one of the TCP flows. One of the symbols $S_i=\{1,2,3\}$ can be
assigned to the $i$th packet loss depending on which TCP lost the
packet. We can consider the sequence of the symbols $\ldots
S_{i-1}S_{i}S_{i+1}\ldots$ generated by the time evolution of our
TCPs. This symbol sequence is a coding of the real congestion window
evolution. Such symbolic coding plays an important role in the
theoretical description of chaotic systems. When an infinite sequence
of symbols codes exactly one or zero real space trajectory the coding
is called \emph{Markov partition}. In this case the symbol sequences
uniquely code the chaotic dynamics. An equivalent definition of the
Markov partition is when each \emph{periodic} symbol sequence codes
exactly one or zero \emph{periodic} orbit of the chaotic system.

In the case of TCP congestion avoidance mode we can demonstrate that
the introduced symbols form a Markov partition. If an infinite
periodic sequence (such as $\ldots122312231223\ldots$) is prescribed
and two different initial congestion window triplets $\bigl((w_1,w_2,w_3)$
and $(w_1',w_2',w_3')\bigr)$ are taken and they evolve according to
the equations (\ref{eq:fluids}), they will reach the loss surface at
different points.  One can prove that the equations (\ref{eq:fluids})
are linearly stable, so they are not capable to increase the
difference between two orbits.  When one of the windows is halved
after the trajectory crosses the loss surface, then the difference
between the orbits is halved in that direction while it remains
unchanged in other directions. The time evolution according to
Eq.~(\ref{eq:fluids}) and the prescribed halvings will decrease the
distance between the two orbits in each period. As all the TCPs should
halve their windows at least once in each cycle the distance between
the two initial triplets is at least halved in each period. This way
we can see that two trajectories started from different initial
conditions converge exponentially to a common periodic orbit.  In the
end we get a unique periodic orbit corresponding to a given periodic
code. So far we forced a given TCP to halve its window according to
the prescribed symbol. In the end we can look at the actual packet
flow generated by the window evolution. The prescribed periodic orbit
can be feasible if the resulting packet flow is in accordance with the
prescribed packet loss sequence or it is not feasible if the resulting
packet flow generates losses in a different order than it has been
assumed.  This way we can decide if the calculated periodic orbit
exists or not.  This procedure ensures that we assign one or zero real
periodic orbits to a periodic sequence and proves the existence of the
Markov partition.
  
\section{The tool-box of chaos}
\label{sec:Toolbox}

The statistical theory of chaos \cite{Ruelle} is based on the symbol
sequences introduced above. If the dynamics is regular (non-chaotic)
then the dynamics is periodic or quasi periodic while the most important
characteristics of chaos is that it endlessly generates topologically
different new trajectories. This is reflected in the way different
systems generate symbol sequences. A length $n$ symbol sequence can
continue in many ways; in average with $a$ number of symbols to form a
length $n+1$ sequence. Thus the number of length $n+1$ sequences $N(n+1)$
can be expressed as
\begin{equation}
  N(n+1)\approx aN(n),
\end{equation}
when the length $n$ is large. In chaotic systems there is more than
one possibility to continue a sequence in average and $a>1$ while in
regular systems $a=1$ for long sequences. Consequently in chaotic
systems the number of possible length $n$ symbols grows exponentially
\begin{equation}
  N(n)\sim a^n\sim e^{\hT n},
\end{equation}  
and the quantity $\hT=\ln(a)>0$ is the \emph{topological entropy}. In
non-chaotic systems the number of sequences grows sub-exponentially
and the topological entropy is zero.

\begin{figure}[htbp]
  \begin{center}
    \psfrag{n}[c][c][1]{$n$}
    \psfrag{N(n)}[c][c][1]{$N(n)$}
    \psfrag{0.95 pm 0.014}[r][r][1]{$\hT=0.953 \pm 0.014$}
    \resizebox{\figwidth}{!}{\includegraphics{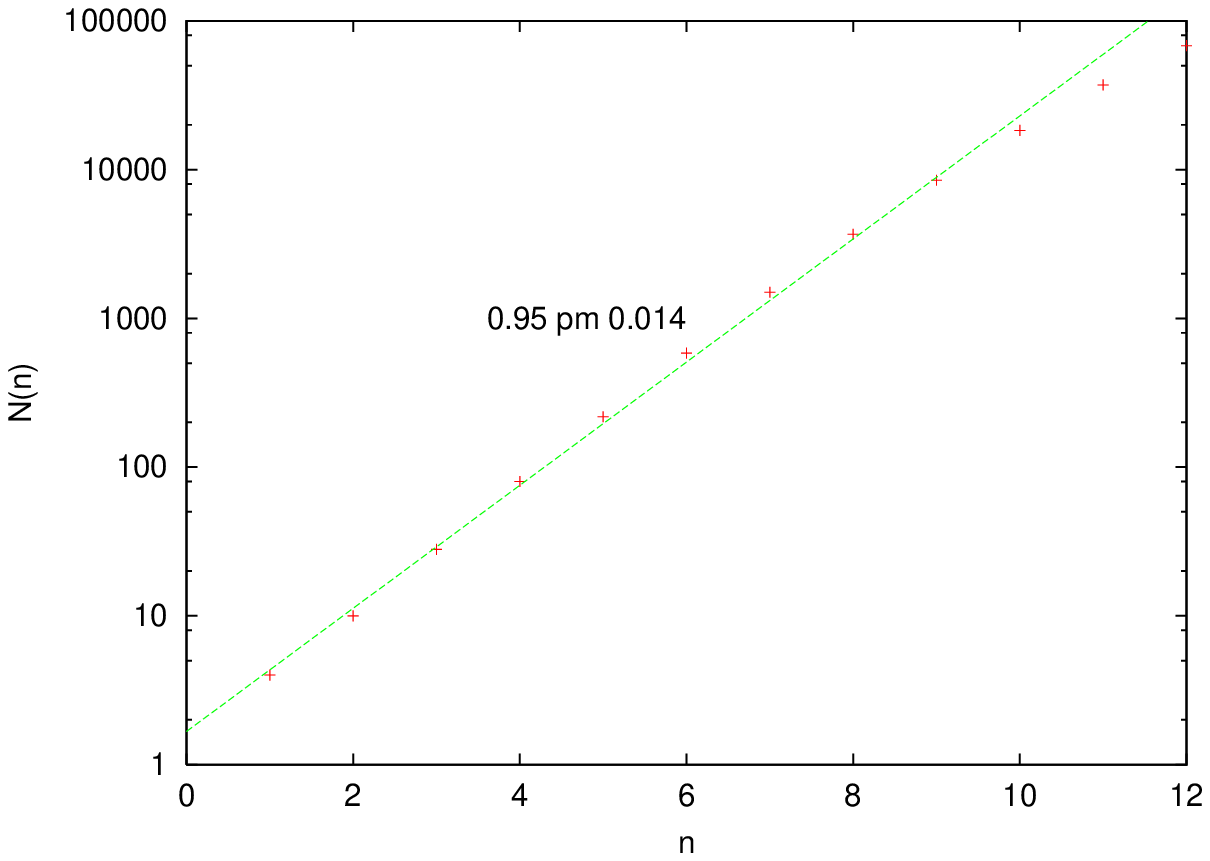}}
    \caption{Number of different symbols as a function of symbol
      length for our system on a semi-logarithmic plot. The fitted line
      is $N(n)=1.669\times2.586^n$}
    \label{fig:topo}
  \end{center}
\end{figure}

To prove that our TCP system really produces chaos we can measure its
topological entropy. In our case we can have a maximum of $3^n$
different symbolic sequences of length $n$. Of course not all of them
are realized by the dynamics since some of them are impossible. For
example infinite sequences consisting of only one or two symbols are
excluded since this would imply that some of the congestion windows
are never halved.  In Fig.~\ref{fig:topo} we show the number of realized
symbol sequences for different lengths $n$ up to $n=12$. The measurement
has been carried out by logging out the symbols (i.e.\ the index of
the TCP which lost a packet) from an \ns\ simulation of the system. We
generated a sequence of 150.000 consecutive symbols and determined how
many different length $n$ sequences exist in it.

We observed that in average $a=2.586$ symbols can follow a given
symbol sequence and the topological entropy is $\hT\approx 0.953$.
This shows that our system is strongly chaotic as the number of
sequences grows with a large exponent, yet it is markedly different
from a stochastic system, where all combination of symbols are allowed
and would result in $a=3$ and a topological entropy of $\ln 3$.

The procedure described so far measures the existence of different
symbol sequences only. We can characterize chaos further by
calculating also the probability $P(S_1,S_2,\ldots,S_n)$ of the
occurrence of the symbol sequence $S_1,S_2,\ldots,S_n$.  In a system
with $L$ symbols ($S_i=\{1,2,\ldots, L\}$) we can visualize this
probability distribution by assigning the number
\begin{equation}
  x=\sum_{i=1}^{n} (S_i-1)L^{-i}
\end{equation}
to each symbol sequence and plotting $P(x)$. In fact $0\leq x < 1$
is the $L$-ary fractional representation of the number represented by
the symbols.

\begin{figure}[htbp]
  \begin{center}
    \psfrag{probability}[c][c][1]{Probability}
    \psfrag{symbols}[c][c][1]{Symbols}
    \resizebox{\figwidth}{!}{\includegraphics{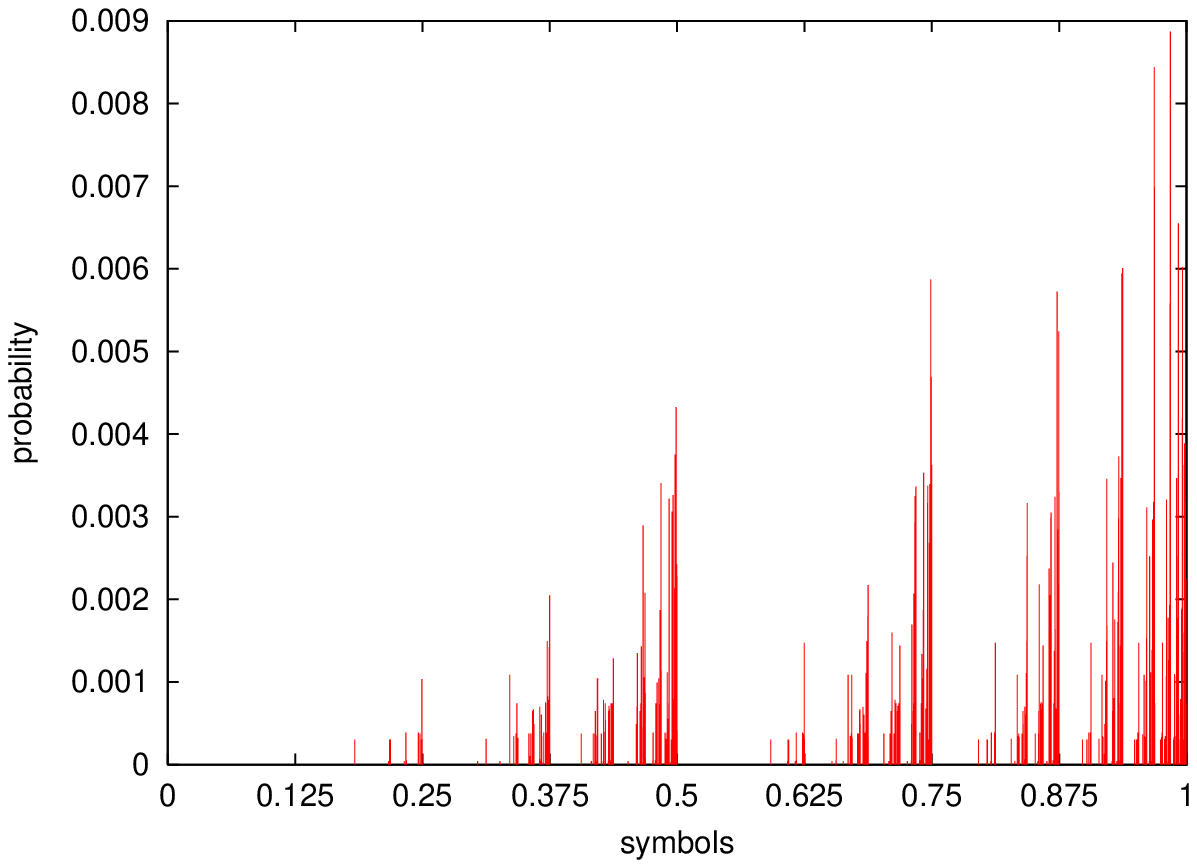}}
    \caption{Probability distribution of symbols of length 10 for our system.}
    \label{fig:prodi}
  \end{center}
\end{figure}

In our system we carried out this analysis and the result is plotted
in Fig.~\ref{fig:prodi}.
It can be clearly seen that the probability distribution is a fractal
in the space of symbols. In fact, in all chaotic systems we should
observe a multifractal distribution and the topological entropy
introduced above is related to the box counting dimension of this
representation. If we cut the $[0,1[$ interval into boxes of size
$\epsilon=1/L^n$ then the number of non-empty boxes is the number of
existing symbols $N(n)$. The box counting dimension is then
\begin{displaymath}
  D_0=\lim_{n\rightarrow \infty} \log N(n)/\log L^n=\hT/\ln L.
\end{displaymath}
In our case then the box counting dimension of the multifractal of
Fig.~\ref{fig:prodi} is 0.864. Note, that this box counting dimension
is \emph{not} related to the box counting dimension of the attractor
discussed before.

Scaling of moments of the probability distribution give further
characterization of the multifractal properties in the symbol space.
We can define the R\'enyi entropies \cite{C_A_M_T_V}:
\begin{equation}
  K_q=\lim_{n\to\infty}\frac 1n\frac{1}{1-q} \ln \sum_{\Sn}
  P^q(S_1,S_2,\ldots,S_n),
\end{equation}
where summation $\Sn$ goes over all possible symbol sequences of
length $n$.  The
quantity $K_q/\ln L$ again measures the $D_q$ generalized dimension of
the multifractal spectrum of the $P(x)$ histogram.  The most important
entropy is the Kolmogorov--Sinai entropy $K_1=\lim_{q\rightarrow
  1}K_q$ which gives the scaling of the Shannon entropy of the
probability distribution:
\begin{align}
  K_1(n)&=-\sum_{\Sn} P(S_1,S_2,\ldots,S_n)
  \ln  P(S_1,S_2,\ldots,S_n),\\
  K_1&=\lim_{n\to\infty}\frac1n K_1(n).
\end{align}

\begin{figure}[htbp]
  \begin{center}
    \psfrag{n}[c][c][1]{$n$}
    \psfrag{Kn}[c][c][1]{$K_1(n)$}
    \psfrag{0.885 pm 0.003}[l][l][1]{$K_1=0.885 \pm 0.003$}
    \resizebox{\figwidth}{!}{\includegraphics{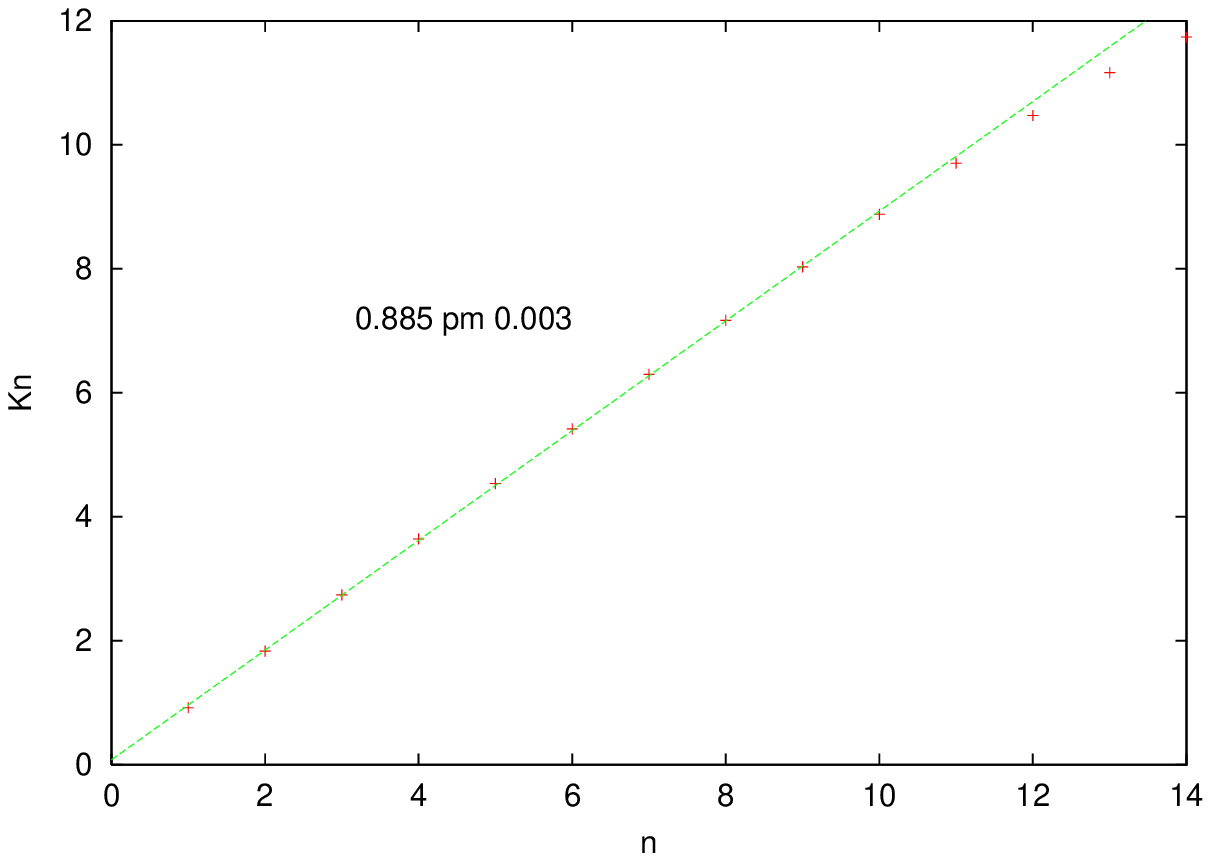}}
    \caption{Kolmogorov--Sinai entropy for our system. 
      $K_1(n)=-\sum_{i=1}^{3^n}p_i\ln p_i=0.885\times n+0.078$}
    \label{fig:kolmo}
  \end{center}
\end{figure}

In Fig.~\ref{fig:kolmo} the Kolmogorov--Sinai entropy is measured for
our system.

The Kolmogorov--Sinai (KS) entropy in a chaotic system is also related
to the Lyapunov exponent $\lambda$ of the mapping of the Poincar\'e
section onto itself. If we consider two nearby trajectories on the
Poincar\'e section---which is the loss surface in our case---then
their initial separation in the phase space $\delta w_0$ grows each
time the trajectories revisit the section. After $n$ revisits the
distance grows exponentially $\delta w_n \approx e^{\lambda n} \delta
w_0$ where the average of the exponent lambda $\langle \lambda
\rangle$ is the Lyapunov exponent of the Poincar\'e section. The KS
entropy gives the Lyapunov exponent $K_1=\langle \lambda \rangle$ in
our system. The positivity of the KS entropy is another indication of
chaos and the exponential sensitivity for the perturbation of
trajectories.

Since the structure of the Markov partition for TCPs is simple even
for larger networks, the topological and KS entropies are easy to
measure in a simulation or can even be determined in a real network.
These quantities measure the complexity of the dynamics and by
evaluating them we can quantify complexity and evaluate TCP and
network models in general.

\section{Cellular chaos}

\begin{figure}[htbp]
  \begin{center}
    \psfrag{basin}[l][l][1]{basin of attraction}
    \psfrag{traj}[c][c][1][15]{trajectory}
    \psfrag{cwnd1}[c][c][1]{$w_1$}
    \psfrag{cwnd2}[c][c][1]{$w_2$}
    \psfrag{cwnd3}[c][c][1]{$w_3$}
    \resizebox{\figwidth}{!}{\includegraphics{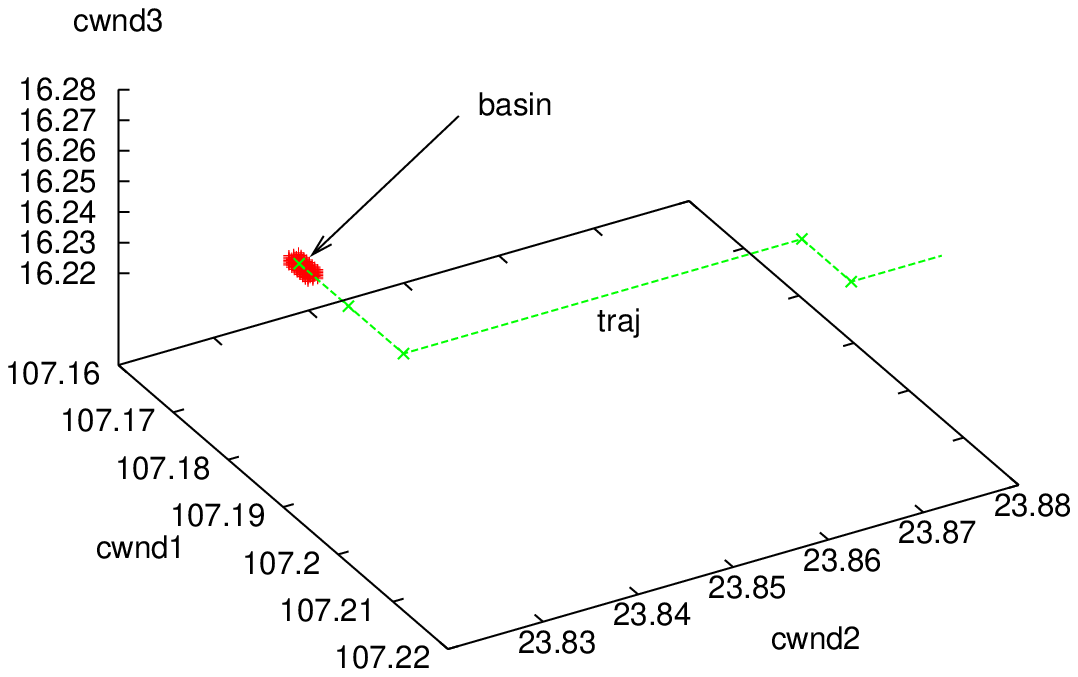}}
    \caption{Our trial TCP trajectory (green) and the points in phase
      space (red) which follow the same trajectory after all three
      windows are halved. The basin of attraction of the starting
      point of the trajectory is approximately a cube of size
      $0.001^3$.}
    \label{fig:ba}
  \end{center}
\end{figure}

The results so far confirmed the hypothesis of chaotic dynamics.
However, the variables of TCP in reality are discrete and not
continuous. We can demonstrate this by applying such a small
perturbation to congestion windows by which we do not change the loss
events happening in the system.  To investigate this we perturbed the
trajectory of Fig.~\ref{subfig:cwnd_nopert} by a small vector $\delta
\mathbf{w}$ and traced the difference between the original and the
perturbed trajectory. We found that there exists a set of
perturbations, shown in Fig.~\ref{fig:ba}, which vanishes after all
windows are halved. The maximal perturbation of this type was found to
be approximately $|\delta\mathbf{w}|<0.001$.
There is a well defined neighborhood around every phase point which
defines the the same behavior of loss dynamics. That is, after all the
congestion windows are halved, the time evolution of congestion
windows becomes identical, mainly because the algorithm sets the
window to its integer part.  This means that the phase space is not
continuous. It is divided into small ``attractive cells'' in which
trajectories converge in finite time. If a trajectory gets into the
cell of another trajectory then they follow the same trajectory later
on.

This property makes TCP chaos very interesting from a theoretical
point of view, since we have a globally chaotic dynamics while the
fine details of the system are non-chaotic. This is not typical in
natural occurrences of chaos but it is potentially important in
engineered systems. The most significant aspect of the cellular
structure is that all trajectories should be periodic. If a trajectory
re-enters a cell visited previously then according to the attractive
nature of the dynamics it will follow the same trajectory again and
will repeat itself.  Since the phase space can be divided into a
finite number of cells a trajectory should at least repeat itself
after visiting all the possible cells. In reality the repetition of a
cell happens much earlier. We can make an estimate of the typical
length and the distribution of the periods of trajectories following
the theory developed by Grebogi, Ott, and Yorke \cite{Grebogi_O_Y} for
chaotic systems with numerical roundoff.

First let us consider the mapping connecting two consecutive points on
the loss surface. Also we can divide the loss surface into $N$
attractive cells. Window values falling into the same cell will follow
the same trajectory later on.  Now suppose that we iterate the mapping
$n$ times, and the orbit visited none of the cells twice.  At this
point we choose a cell random from the $n$ visited one, and try to
calculate the probability that in step $n+1$ the system visits the
chosen cell. The following calculations assume that the dynamics of
the map is mixing (see \cite{Grebogi_O_Y} for details). The result of
mixing is that the typical value of $n-j$ is large, where $j$ is the
iterate when the chosen cell was first visited. Thus $\bar l$ becomes
large, where $l=n-j+1$, and bar denotes expectation value.  Now the
probability of repeating the chosen cell in step $n+1$ is equal to the
probability of visiting that cell. On the other hand, due to random
initial conditions, the probability that a given cell is visited in
step $j$ is also equal to the probability of the attractor to visit
that cell. Combining these, for the probability of repeating the cell
in step $n+1$ that was first visited in step $j$ one obtains
\begin{equation}
  \label{eq:avg_p}
  \langle p \rangle = \sum_i p_i^2,
\end{equation}
where $p_i$ is the probability with which the orbit visits the $i$th
cell (i.e.\ the measure of the attractor in that cell). So the
probability of repeating any of the previous cells in step $n+1$,
supposing that the orbit did not repeat during the previous $n$ steps
is:
\begin{equation}
  p_{\text{rep}}(n)=n\langle p\rangle.
\end{equation}
Thus the probability that an orbit of length $n$ has no repeats is
\begin{equation}
  p_{\text{norep}}(n) =\prod_{k=1}^{n-1} (1-p_{\text{rep}} (k))
  = \prod_{k=1}^{n-1} (1- k\langle p \rangle ).
\end{equation}
One can approximate $\ln p_{\text{norep}}(n)$ for $n\langle p\rangle
\ll 1$:
\begin{equation}
  \begin{split}
    \ln p_{\text{norep}} (n)&=\sum_{k=1}^{n-1} \ln (1-k\langle
    p\rangle )\\ 
    &\approx - \sum_{k=1}^{n-1}k \langle p\rangle\approx
    - n^2 \langle p\rangle /2,
  \end{split}
\end{equation}
which yields
\begin{equation}
  p_{\text{norep}}(n)\approx \exp [ -(n^2 \langle p\rangle /2)].
\end{equation}
The condition for this approximation ($n\langle p\rangle \ll 1$) is
justified if $\bar l \langle p \rangle \ll 1$. As we shall see, $\bar
l \sim \langle p \rangle ^{1/2}$, thus the above condition is
equivalent to $\bar l \gg 1$. The probability that an orbit goes $n$
steps without repeat and then repeats on step $(n+1)$ is
\begin{equation}
  p(n)=p_{\text{rep}}(n)\,p_{\text{norep}}(n).
\end{equation}
If we assume that the probability of returning to any of the preceding
cells is equal, then if the first repeat occurs at step $n+1$, the
probability that $n-j+1=l$ is
\begin{equation}
  p_{\text{p}} (n,l)=
  \left\{
    \begin{array}{ll}
      1/n & \text{for $l\le n$,}\\
      0 & \text{for $l>n$}.
    \end{array}
  \right.
\end{equation}
For large $l$ the probability density of $l$ becomes
\begin{equation}
  \begin{split}
    P(l)&=\int_0^{\infty} p(n)\, p_{\text{p}}(n,l)\, dn\\
    &=\frac{\sqrt{8}}{\pi} \langle p
    \rangle^{1/2} F\left(\langle p\rangle ^{1/2} l\right),
  \end{split}
\end{equation}
where we substituted the results above, and where
\begin{displaymath}
  F(y)=\sqrt{\pi / 8}\int_y^{\infty}\exp(-x^2 /2)\,dx.
\end{displaymath}
Calculating the expectation value of $l$, we obtain
the average length of periodic orbits generated by the discrete,
cellular structure of the phase space:
\begin{equation}
  \label{eq:avg_q}
  \bar l= \int_0^{\infty} l\, P(l)\, dl = 
  \sqrt{\pi /8}\,\langle p \rangle^{-1/2}.
\end{equation}
In the next section we investigate the implications of the periodicity
of the trajectories.

\section{Exploring periodic orbits}

Until now three parallel TCPs were studied in our simulation scenario.
In such situation the number of cells which can be distinguished from
each other in the phase space is in the order of $\sim10^{15}$. Also,
the length of a typical periodic orbit is enormous and we do not
expect to observe them in realistic simulations.

Therefore, it seems more reasonable to find periodic orbits when only
two TCPs are operating in the network. For this scenario we used the
same network model that we have introduced in Fig.~\ref{fig:model}
except that the TCP with the largest round-trip time was removed and
the buffer size was set to $B=50$.

\begin{figure}[tbhp]
  \begin{center}
    \psfrag{w_1}[c][c][1]{$w_1$}
    \psfrag{w_2}[c][c][1]{$w_2$}
    \resizebox{\figwidth}{!}{\includegraphics{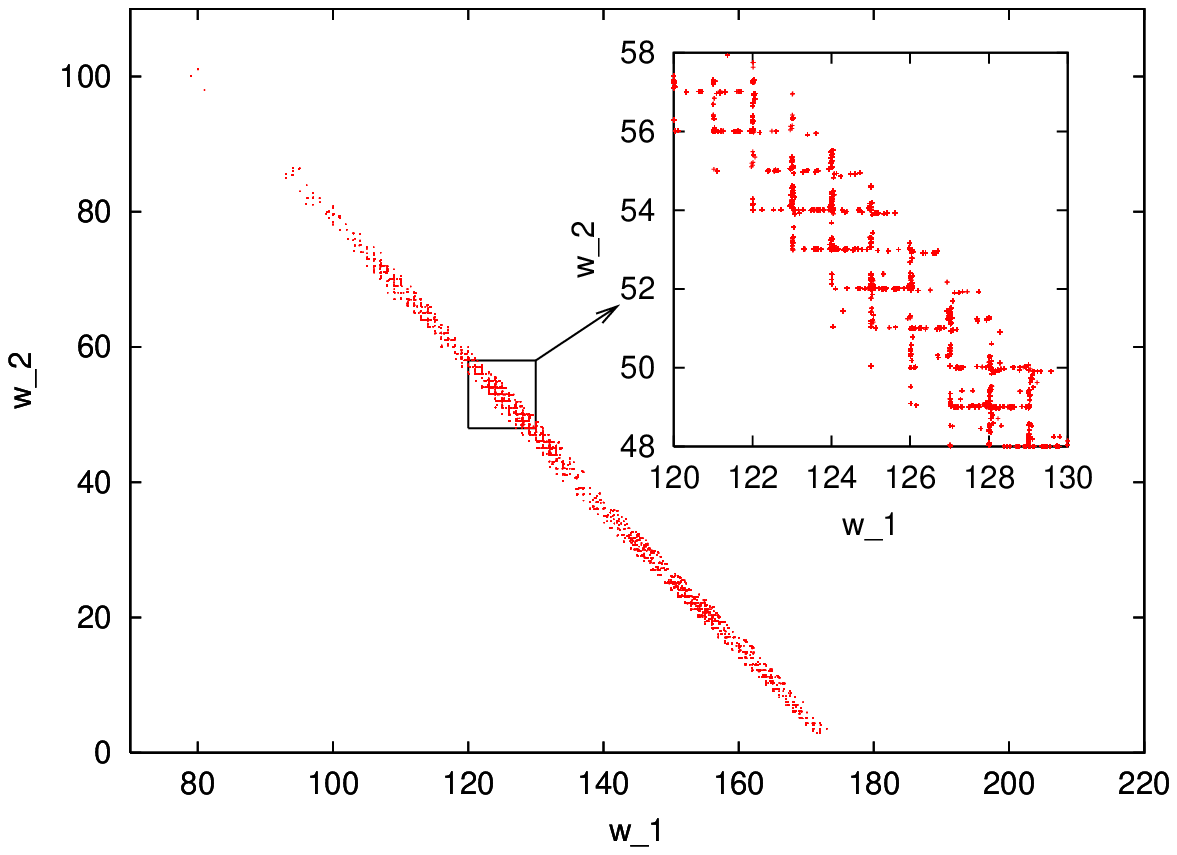}}
    \caption{Poincar\'e section of the phase space. Congestion window
      values for the two TCPs at packet loss times. A small part of
      the Poincar\'e section is zoomed to demonstrate the fine
      structure of the section.}
    \label{fig:pss_2d}
  \end{center}
\end{figure}

Applying the tools developed in
Sections~\ref{sec:Characterizing}--\ref{sec:Toolbox} we can study the
chaotic properties of this system. First the Poincar\'e surface of
section is investigated (Fig.~\ref{fig:pss_2d}). Similarly to the
three dimensional case, the sum of the congestion windows has to
approximately satisfy the $\sum_i w_i=Q$ condition at packet loss
times, where $Q$ is the maximum number of packets on the lines and in
the buffer. This condition defines the ``line of loss'' inside the two
dimensional window phase space now. The detailed structure of the
Poincar\'e section is shown in the inset in Fig.~\ref{fig:pss_2d}. It
can be seen that the Poincar\'e section is not exactly a line but a
narrow grid spreading around the ideal line.
This shape can be explained by the packet drop mechanism at the
bottleneck buffer. A detailed discussion of this mechanism is given in
Section~\ref{sec:Toward}.

\begin{figure}[tbhp]
  \begin{center}
    \psfrag{n}[c][c][1]{$n$}
    \psfrag{N(n)}[c][c][1]{$N(n)$}
    \psfrag{0.54 pm 0.013}[l][l][1]{$\hT=0.54 \pm 0.013$}
    \psfrag{606.3}[l][l][1]{$606.3$}
    \resizebox{\figwidth}{!}{\includegraphics{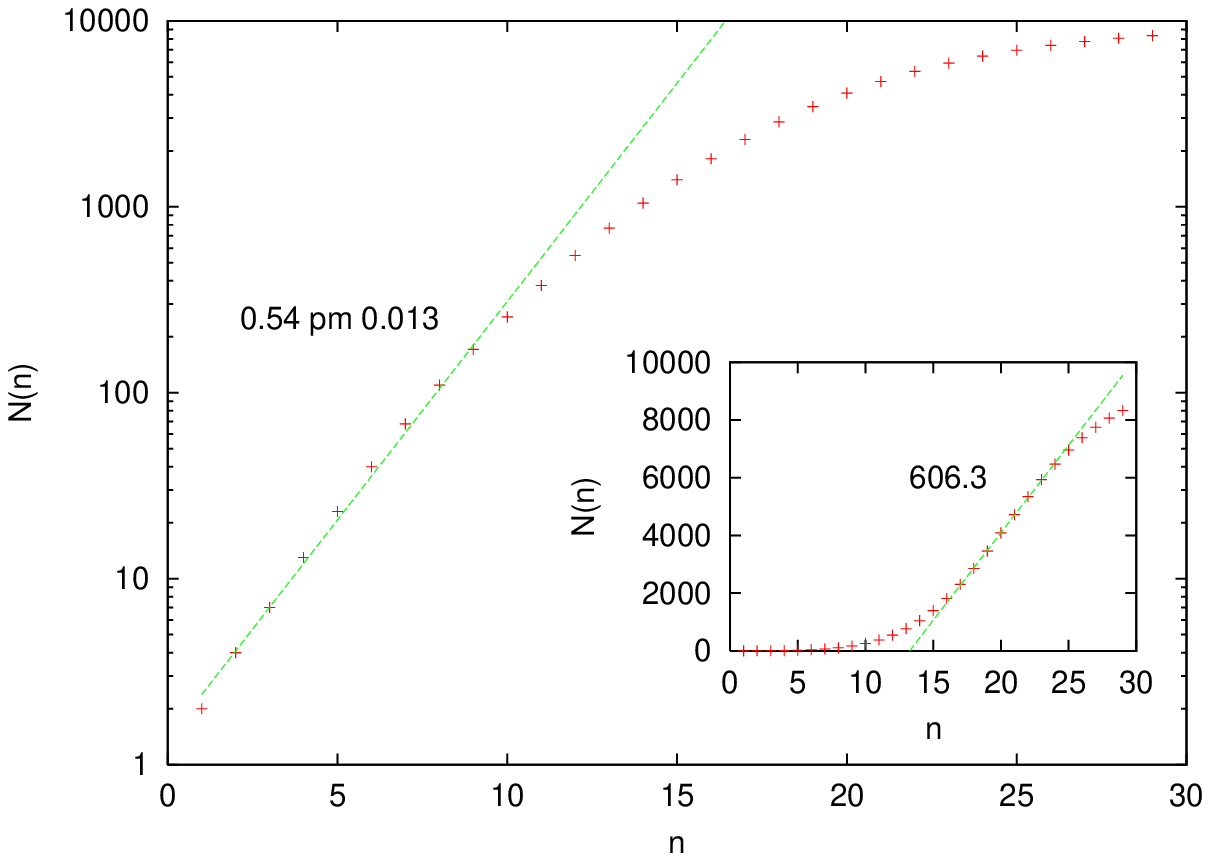}}
    \caption{Number of different symbols as the function of symbol
      length $n$ for 2 TCP on a semi-logarithmic and on a normal plot in
      the inset. The fitted lines are 
      $N(n)=1.387\times1.718^n$ and $N(n)=606.3\times n-8034$.}
    \label{fig:topo_2d}
  \end{center}
\end{figure}

\begin{figure}[tbhp]
  \begin{center}
    \psfrag{n}[c][c][1]{$n$}
    \psfrag{Kn}[c][c][1]{$K_1(n)$}
    \psfrag{0.44 pm 0.002}[l][l][1]{$K_1=0.440 \pm 0.002$}
    \resizebox{\figwidth}{!}{\includegraphics{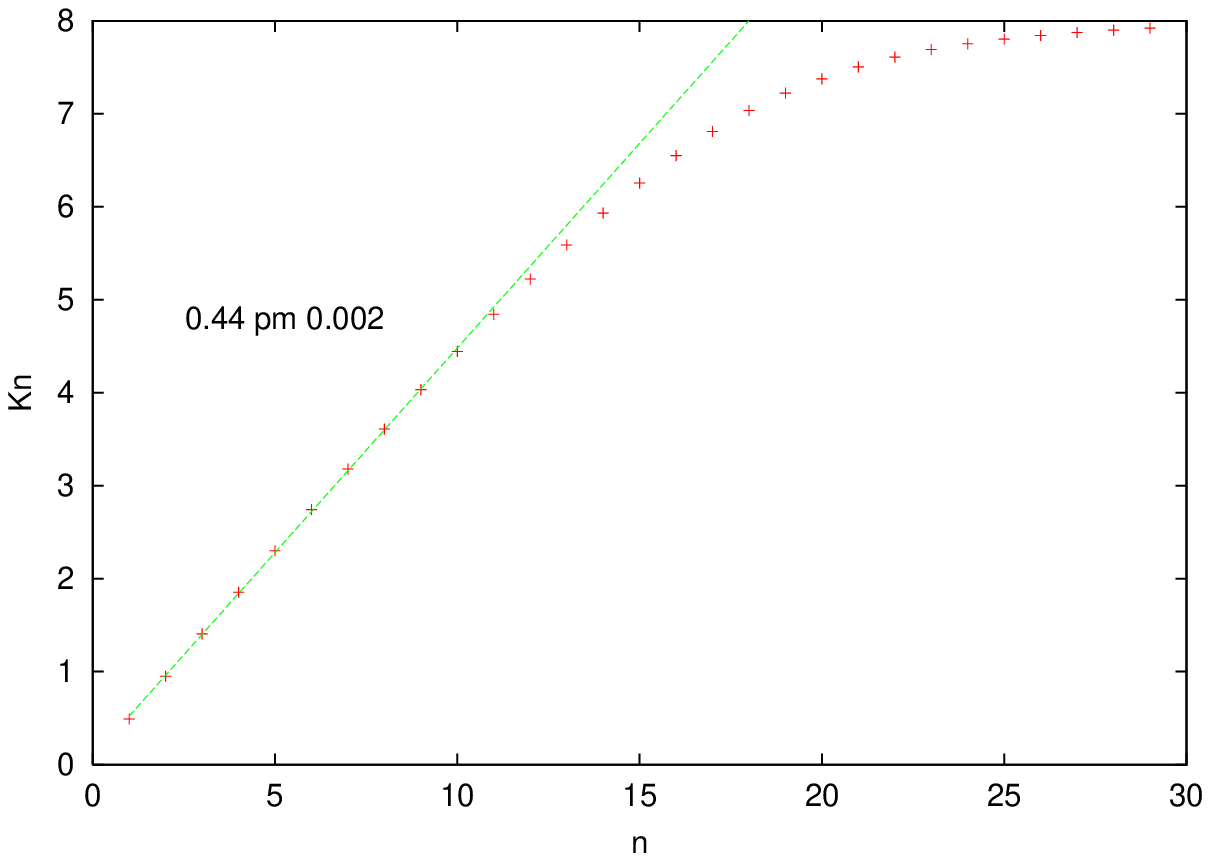}}
    \caption{Kolmogorov entropy for 2 TCP scenario. 
      $K_1(n)=-\sum_{i=1}^{2^n}p_i\ln p_i=0.440\times n+0.081$}
    \label{fig:kolm_2d}
  \end{center}
\end{figure}

To measure the complexity of the TCP dynamics in this situation we
apply the symbolic coding that was introduced in
Section~\ref{sec:Symbols}. Then, the topological
(Fig.~\ref{fig:topo_2d}) and the Kolmogorov--Sinai
(Fig.~\ref{fig:kolm_2d}) entropies are estimated. For topological
entropy $\hT=0.54\pm0.01$, while for the KS entropy $K_1=0.440\pm0.002$
have been obtained. Both values indicate strong mixing and the presence of
chaos in the system. However, the maximum number of states is limited
due to the discrete phase space, and TCP must enter into a periodic
cycle after an initial transient period. If the orbit realized in the
simulation is in fact periodic with some large period, then
the number of existing length $n$ sequences $N(n)$
increases only linearly with $n$ if $n$ is sufficiently large. In case of
two TCP this linear growth is observable above $n=15$ (see the inset in
Fig.~\ref{fig:topo_2d}). Also the entropy $K_1(n)$ saturates above $n=15$
and the KS entropy $K_1$ goes to zero indicating that the long time
behavior of the system is periodic.

In our 2-TCP scenario periodic orbits were identified.  For the given
simulation setup the length of the period was found to be independent
of the initial conditions. The length of the period was $l=1744$.
Using this and Eqs.~(\ref{eq:avg_p}) and (\ref{eq:avg_q}) we can
estimate the order of magnitude of the number of cells in this system.
The cells can be indexed by the corresponding symbol
sequences. Suppose that we can index uniquely all the cells with
symbol sequences of length $n^*$. In this case the number of cells is
approximately given by $e^{K_0n^*}$. The probability of repeating a
cell is then given by
\begin{equation}
  \label{eq:avg_p2}
  \langle p\rangle=
  \sum_{\left\{S\right\}_{n^*}}P^2(S_1,S_2,\ldots,S_{n^*})
  \approx e^{-K_2 n^*},
\end{equation}
where $K_2$ is the R\'enyi entropy for $q=2$. Combining
(\ref{eq:avg_q}), the estimate for the number of cells $e^{K_0n^*}$
and (\ref{eq:avg_p2}) one obtains
\begin{equation}
  \label{eq:q_estimate}
  N\approx e^{K_0 n^*}\approx\langle p\rangle^{-K_0/K_2}.
\end{equation}
On the other hand (\ref{eq:avg_q}) implies
\begin{equation}
  \label{eq:avg_p3}
  \langle p\rangle=\frac{\pi}8\frac1{\bar{l}^2},
\end{equation}
and finally:
\begin{equation}
  \label{eq:ize}
  N\approx \left(\frac8{\pi}\bar{l}^2\right)^{K_0/K_2}.
\end{equation}

Assuming that the obtained period $l=1744$ is a good approximation of
the average period and using the values $K_0=0.54$ and $K_2=0.39$ we
get $N\approx3\cdot10^9$. This is in accordance with expectations,
since the area in the phase space visited by the congestion window
trajectories is approximately a triangle with area $90\times90/2$ as
one can see in Fig.~\ref{fig:pss_2d}, while the area of a cell is
approximately $0.001^2$ in accordance with the findings of
Fig.~\ref{fig:ba}. This implies that the visited part of the phase
space contains approximately $4\cdot10^9$ cells.

\section{Toward chaos aware modeling}
\label{sec:Toward}

In the previous sections we managed to collect all the basic elements
of chaos in TCP congestion avoidance.  Now we would like to show how
we can integrate all these into a model without stochastic elements
which is able to reproduce all the main features.

The time evolution of congestion windows consists of two parts. The
first one is when windows grow steadily.  This part is
sufficiently well described by existing fluid models like the one
given by Eq. (\ref{eq:fluids}). In our case just for qualitative
comparison we can use a version, where we assume that the buffer is
non-empty:
\begin{equation}
  \frac{d w_i}{dt}=\frac{1}{T_0+T_i+bP/C_0},
  \label{eq:diff}
\end{equation}
where $b$ is the actual number of packets in the buffer.  To keep
things simple, we estimate the number of packets in the system by the
sum of congestion windows and the packets traveling in the lines by
$C_0T_0/P+\frac{1}{3}C_0(T_1+T_2+T_3)/P$. This way the queue length in
the buffer is estimated as
\begin{displaymath}
  b=\sum_i w_i - C_0T_0/P+\frac{1}{3}C_0(T_1+T_2+T_3)/P.
\end{displaymath}
  
The second and more crucial part of the evolution is the dynamics at
packet loss events. In our model we can simplify matters and say that
a packet loss occurs whenever the buffer is full and $b=B$ holds. At
that time we halve and take the integer part of the congestion window
of the TCP whose packet has been lost $w_i'=[w_i/2]$, where $[x]$
denotes the integer part of $x$.

The elements of the model discussed so far do not generate chaotic
dynamics.  Now, the really important new element is how we make
decision on which TCP loses a packet at a given buffer overflow.  In
traditional stochastic models this step is random. The argument behind
this is that the details of the packet flow are difficult to model on
a macroscopic scale and if the packets are well mixed (which is a
realistic assumption) then the chance of packet loss for each TCP is
proportional with the rate a given TCP sends packets into the buffer.
However in reality the packet loss process is more deterministic and
predictable than it might seem at first sight.  A packet is shifted
out of the buffer at each time unit $\tau=P/C$, where $P$ is the
packet size and $C$ is the bandwidth of outgoing packets. When the
buffer is full then packet loss will not occur until only one packet
comes in within the time slot $\tau$. Packet loss does happen when
more than one packets arrive in such a critical time slot.  In
congestion avoidance mode this can happen if two or more TCP send in
packets within the critical slot or if some of the TCPs sends in two
packets. Single packets are sent in when a TCP receives an
acknowledgement packet, double packets are sent in when in addition
the congestion window crosses an integer number. If there are more
than one packets sent in during the critical time slot then the one
which will make it to the buffer is selected by the exact timing of
the packets. This phenomena is called \emph{phase effect} and has been
discovered by Floyd and Jacobson in Ref.~\cite{Floyd_J}. Depending on
the concrete setup one can determine which TCP will be the winner out
of the ones sending packets in the same slot.

In our case the phase analysis shows that the winner TCP is
accidentally the one with the largest delay. Consequently this TCP
will win all the battles for the critical last buffer space. The only
way this TCP can lose packets is when it sends in two packets. This is
the reason behind the dominance of this TCP and its very regular
behavior observable in Fig.~\ref{subfig:cwnd_nopert}.

It is clear now that phase effects should be integrated into our
model. The difficulty is that the fluid congestion window model is not
able to predict when packets are sent out by TCPs. On the other hand,
if we start modeling the concrete packet process of TCP then we are
back at packet level modeling. So, we have to make a good compromise
which saves the simplicity of the fluid equations while keeps the key
elements of the packet loss process including phase effects. This can
be achieved if we attribute a packet process to the macroscopic fluid
evolution of congestion windows.  This can be done as follows. Each
time the TCP sends out a packet its congestion window is increased by
$1/[w]$.  Such a way in each round trip time the window increases with
$1$ and $[w]$ number of packets are sent out. In our simulation we can
reverse this and we can calculate the packets sent out by the TCP from
the actual macroscopic window changes. We can calculate the fluid
window values using the differential equations (\ref{eq:diff}) at the
beginning of each time slot:
\begin{displaymath}
  w_i(n)=w_i(t=n\cdot\tau).
\end{displaymath}
We can compare $w_i(n+1)$ and $w_i(n)$ and see if the window crosses
in this time slot an integer multiple of $1/[w]$.  (This can be done
by checking if $[w_i(n+1)[w_i(n)]]$ and $[w_i(n)[w_i(n)]]$ differ or
not.)  If it does then we assume that a packet has been sent out by
the TCP. If the window crosses also an integer then we assume that two
packets has been sent out in that time slot. This attributed packet
process is relevant when the buffer becomes full and the windows reach
the loss surface. At this point we should check the sending status of
TCPs.  If more than one packets arrive during the time slot we let the
winner TCP to fill the buffer space while the rest of the TCPs loose
packets and halve their windows.

\begin{figure}[h!tbp]
  \begin{center}
    \psfrag{w_1}[c][c][1]{$w_1$}
    \psfrag{w_2}[c][c][1]{$w_2$}
    \psfrag{w_3}[c][c][1]{$w_3$}
    \resizebox{\figwidth}{!}{\includegraphics{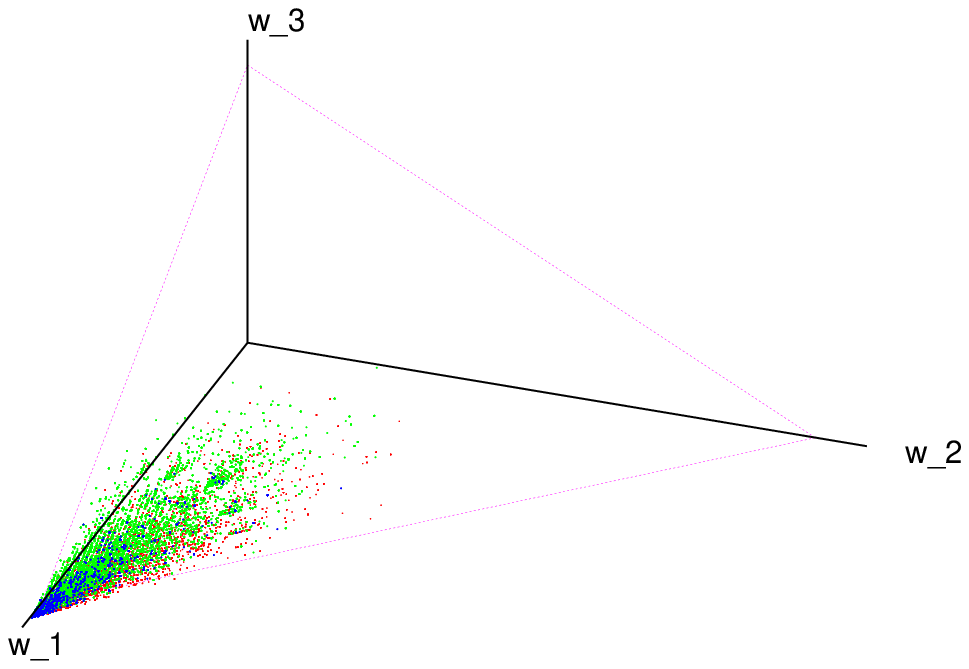}}
    \caption{Poincar\'e section of the phase space for the model
      described in Section~\ref{sec:Toward}. Congestion window
      values for the three TCPs at packet loss times. Colours depend
      on which TCP lost a packet. The depicted (pink) surface is where
      the sum of congestion windows is approximately equal with the
      maximum number of packets which can travel on the lines or stay
      in the buffer.}
    \label{fig:pss_model}
  \end{center}
\end{figure}

We carried out such a simulation for our model system with three TCP
flows. The results are in accordance with the results obtained from
the packet level ns simulation. In Fig.~\ref{fig:pss_model} we can see
that a loss surface similar to Fig.~\ref{fig:pss} is formed.

\begin{figure}[htbp]
  \begin{center}
    \psfrag{log10(l)}[c][c][1]{$\log_{10}(\epsilon)$}
    \psfrag{log10(N(l))}[c][c][1]{$\log_{10}(N(\epsilon))$}
    \psfrag{log10(sum(p_i^2))}[c][c][1]{$\log_{10}(\sum p_i^2)$}
    \psfrag{1.49 pm 0.03}[l][l][1]{$D_0=1.49\pm 0.03$}
    \psfrag{1.46 pm 0.07}[r][r][1]{$1.46\pm 0.07$}
    \resizebox{\figwidth}{!}{\includegraphics{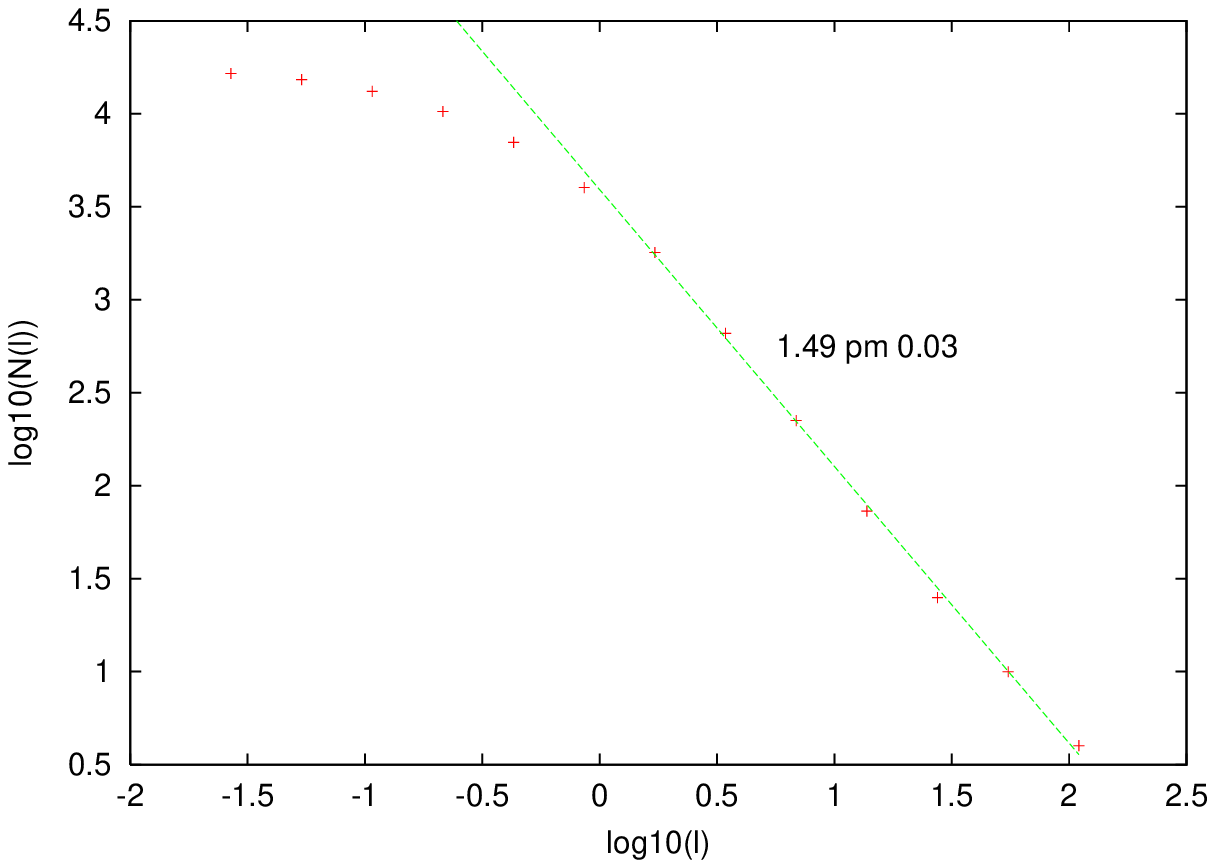}}
    \caption{Fractal dimension of the projection of loss events in the
      phase space obtained from the model.}
    \label{fig:fracdim_model}
  \end{center}
\end{figure}

The fractal nature of the Poincar\'e section is also preserved.
The fractal dimension of the projection is shown in
Fig.~\ref{fig:fracdim_model}. 

\begin{figure}[tbhp]
  \begin{center}
    \psfrag{n}[c][c][1]{$n$}
    \psfrag{N(n)}[c][c][1]{$N(n)$}
    \psfrag{0.862 pm 0.023}[l][l][1]{$K_0=0.862 \pm 0.023$}
    \resizebox{\figwidth}{!}{\includegraphics{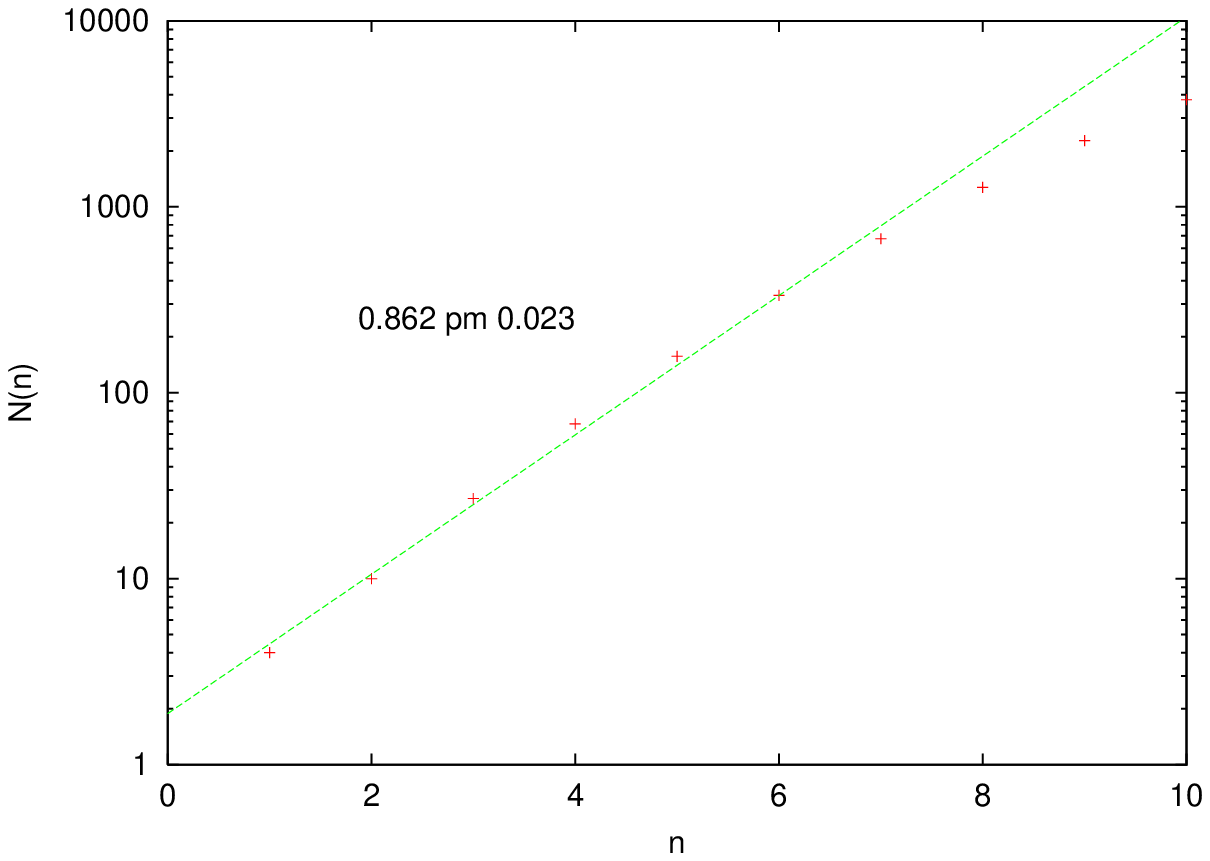}}
    \caption{Topological entropy for 3 TCP obtained from the model.}
    \label{fig:topo_model}
  \end{center}
\end{figure}
\begin{figure}[htbp]
  \begin{center}
    \psfrag{n}[c][c][1]{$n$}
    \psfrag{Kn}[c][c][1]{$K_1(n)$}
    \psfrag{0.753 pm 0.006}[l][l][1]{$K_1=0.753 \pm 0.006$}
    \resizebox{\figwidth}{!}{\includegraphics{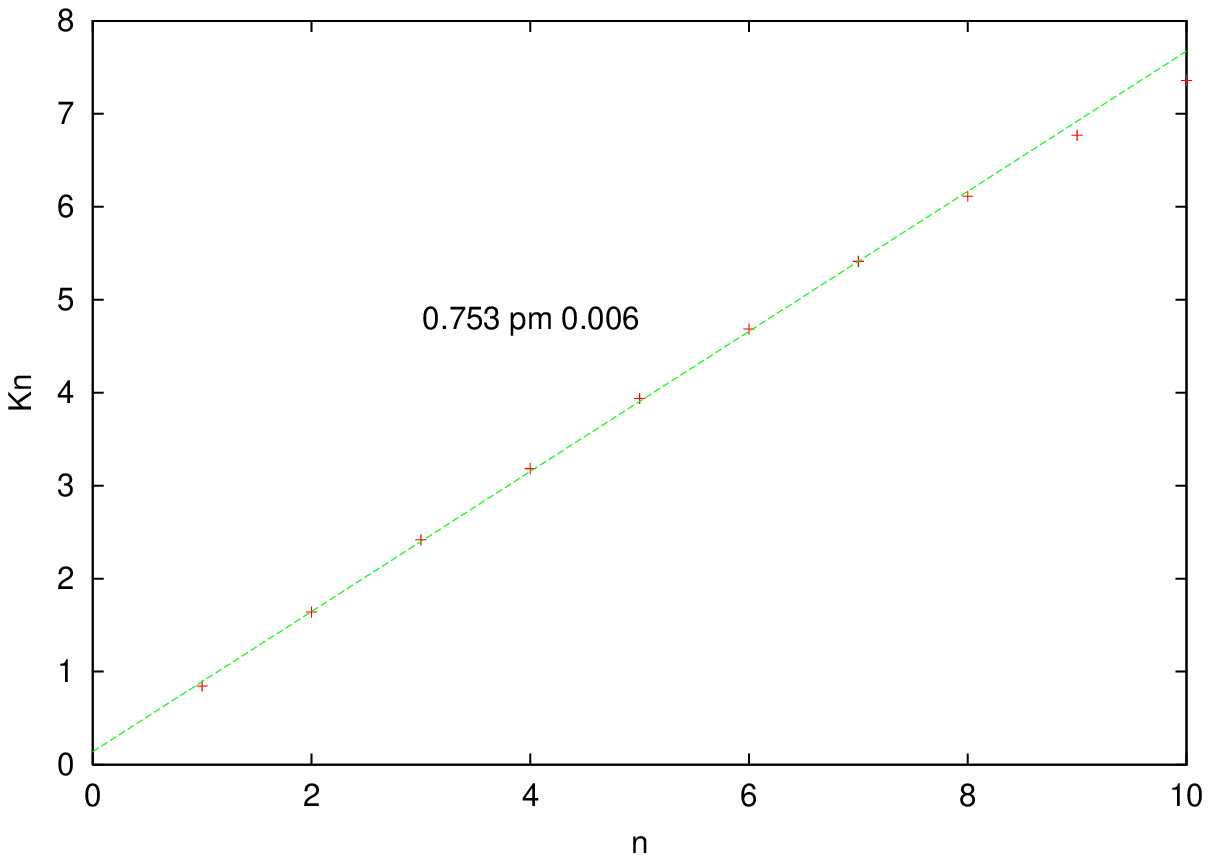}}
    \caption{Kolmogorov--Sinai entropy for 3 TCP obtained from the
      model. $K_n=-\sum_{i=1}^{3^n} p_i\ln p_i=0.753\times n+0.108$}
    \label{fig:kolm_model}
  \end{center}
\end{figure}

The model creates non-trivial symbolic dynamics similar to the one
observed in the \ns\ simulation. The measured topological and KS
entropies are shown in Figs.~\ref{fig:topo_model} and
\ref{fig:kolm_model}.

This model can be further improved and extended by a more detailed
model of the packet loss process including the time slots before the
buffer reaches its critical state. The three TCPs are able to send in
a maximum of 6 packets during a single time slot and an overflow can
happen in a buffer with 5 free packet slots. This effect can cause the
broadening of the loss surface as in Fig.~\ref{fig:pss_2d}.  Also the
time elapsed between the packet loss event and its notification can be
introduced and a more accurate description of the relationship between
the round trip time and window values is needed to obtain
quantitatively better results.

\section{Conclusions}

In this paper we investigated the chaotic properties of TCPs operating
in congestion avoidance mode. We showed that chaotic behavior is
general even in the case of low packet loss probability.  We
demonstrated that the dynamics can be viewed as a smooth time
evolution between packet losses and the relevant features of chaos
might be described by the investigation of the Poincar\'e section
defined by packet loss events. Chaotic dynamics can be characterized
by symbol sequences and we introduced topological and
Kolmogorov--Sinai entropies, whose values confirmed the hypothesis of
chaos.  Due to the deterministic nature of the system, in contrast
to stochastic models, not all possible symbol sequences are
realized, some of them are excluded by the dynamics. Accordingly the
topological entropy is significantly less than its possible maximal
value $\ln L$. The positive Kolmogorov--Sinai entropy indicates that
the Lyapunov exponent of the system is positive and that the
distribution of symbol sequence probabilities is multifractal.  We
also proved that due to the cellular structure of the phase space the
long time behavior of the system is inherently periodic.

To reproduce these phenomena, we introduced a simple model that
incorporates a fluid model of congestion window evolution but also
handles packet losses in a deterministic way.  In spite of the
extensive simplifications we used in the model the basic dynamical
characteristics of the TCP dynamics such as the fractal dimension of
the attractor and various entropies came out in a qualitatively
correct form. The model is also able to count for the apparent
unfairness of TCP flows.  What we demonstrated here is that instead of
a stochastic model of packet loss events, a deterministic model can be
worked out and combined with the macroscopic fluid equations. This new
model is able to reproduce important phase effects.  While the
attributed packet generation is not fully identical with the packet
generation of a packet level simulation, yet it generates packet flows
whose statistical properties are close to the real situation. The
model correctly reproduces the loss probabilities of packets arriving
in near coincidence into the buffer.  This property seems to be
essential in generating a proper chaotic time evolution.

\section*{Acknowledgment}

The authors would like to thank for the fruitful discussions with
M.~Boda, A.~Veres, J.~St\'eger and S.~Moln\'ar. G.~V. thanks the
support of the Hungarian Scientific Research Fund (OTKA T032437 and
T037903).

\bibliographystyle{IEEE} 
\bibliography{chaos}

\begin{thebibliography}{10}

\bibitem{Veres_B}
A.~Veres and M.~Boda,
\newblock ``The chaotic nature of {TCP} congestion control,''
\newblock in {\em IEEE INFOCOM'2000}, March 2000.

\bibitem{Veres_K_M_V}
A.~Veres, Zs. Kenesi, S.~Moln\'ar, and G.~Vattay,
\newblock ``On the propagation of long-range dependence in the {Internet},''
\newblock in {\em ACM SIGCOMM 2000}, Stockholm, Sweden, August 2000.

\bibitem{Guo_C_M}
L.~Guo, M.~Crovella, and I.~Matta,
\newblock ``{TCP} congestion control and heavy tails,''
\newblock Tech. Rep. BUCS-TR-2000-017, Computer Science Dep., Boston
  University, 2000.

\bibitem{Figue_L_M_T}
D.~R. Figueiredo, B.~Liu, V.~Mishra, and D.~Towsley,
\newblock ``On the autocorrelation structure of {TCP} traffic,''
\newblock Tech. Rep. 00-55, Dep. of Computer Science, University of
  Massachusetts, Amherst, November 2000.

\bibitem{Fekete_V}
A.~Fekete and G.~Vattay,
\newblock ``Self-similarity in bottleneck buffers,''
\newblock in {\em Proceedings of Globecom 2001}, December 2001.

\bibitem{RFC2001}
W.~Stevens,
\newblock {\em {TCP} Slow Start, Congestion Avoidance, Fast Retransmit and Fast
  Recovery Algorithms},
\newblock RFC 2001, Jan. 1997.

\bibitem{ns}
``{UCB}/{LBNL}/{VINT} network simulator -- ns (version 2),''
  http://www-mash.cs.berkeley.edu/ns/.

\bibitem{Zhang_Q_K}
Y.~Zhang, L.~Qiu, and S.~Keshav,
\newblock ``Understanding the performance of many {TCP} flows,''
\newblock {\em Computer Networks}, vol. 37, pp. 277--306, 2001.

\bibitem{Vicsek}
T.~Vicsek,
\newblock {\em Fractal {G}rowth {P}henomena},
\newblock World Scientific, Singapore, 2nd edition, 1992.

\bibitem{Ott_K_M}
T.~J. Ott, J.~H.~B. Kemperman, and M.~Mathis,
\newblock ``The stationary behavior of ideal {TCP} congestion avoidance,''
\newblock in {\em Proceedings of IEEE INFOCOM'99}, New York, 1999.

\bibitem{Misra_O}
Archan Misra and Teunis Ott,
\newblock ``The window distribution of idealized {TCP} congestion avoidance
  with variable packet loss,''
\newblock in {\em INFOCOM'99}, March 1999.

\bibitem{Ruelle}
D.~Ruelle,
\newblock {\em Statistical {M}echanics, {T}hermodynamic {F}ormalism},
\newblock Addison-Wesley, Reading MA, 1978.

\bibitem{C_A_M_T_V}
{\em Detailed discussion of various entropies characterizing chaotic
systems can be found in:} 
P.~Cvitanovi\'c, R.~Artuso, R.~Mainieri, G.~Tanner, and G.~Vattay,
\newblock {\em Classical and Quantum Chaos},
\newblock Niels Bohr Institute, Copenhagen, 2001,
\newblock http://www.nbi.dk/ChaosBook/

\bibitem{Grebogi_O_Y}
C.~Grebogi, E.~Ott, and J.~A. Yorke,
\newblock ``Roundoff-induced periodicity and the correlation dimension of
  chaotic attractors,''
\newblock {\em Physical Review A}, vol. 38, no. 7, pp. 3688--3692, October
  1988.

\bibitem{Floyd_J}
S.~Floyd and V.~Jacobson,
\newblock ``On traffic phase effects in packet-switched gateways,''
\newblock {\em Internetworking: Research and Experience}, vol. 3, pp. 115--156,
  1992.

\end{thebibliography}

\end{document}